\documentclass[sn-mathphys-num, referee]{sn-jnl}

\usepackage[ngerman,english]{babel} %
\usepackage[utf8]{inputenc}

\usepackage{xcolor,soul}

\usepackage{amsmath,amssymb, mathtools} %

\usepackage{xargs}                      %

\usepackage{nicefrac}
\usepackage{braket}
\usepackage[mathscr]{eucal}	
\usepackage{bm}
\usepackage{siunitx}
\usepackage{graphicx}%
\usepackage{multirow}%
\usepackage{amsmath,amssymb,amsfonts}%
\usepackage{pdfcomment}
\usepackage{subfig}
\usepackage[nohyperlinks, nolist]{acronym}

\renewcommand{\vec}{\mathbf}

\unnumbered%

\begin{document}

\title[Magnon momentum microscopy]{Soft-X-ray momentum microscopy of nonlinear magnon interactions below 100-nm wavelength} %

\author*[1, 2]{Steffen Wittrock}\email{steffen.wittrock@helmholtz-berlin.de}
\author[1]{Christopher Klose}
\author[3,2]{Salvatore Perna}
\author[4]{Korbinian Baumgaertl}
\author[4]{Andrea Mucchietto}
\author[1]{Michael Schneider}
\author[1]{Josefin Fuchs}
\author[1]{Victor Deinhart}
\author[2]{Tamer Karaman}
\author[4,5]{Dirk Grundler}
\author[1,6]{Stefan Eisebitt}
\author*[1]{Bastian Pfau}\email{pfau@mbi-berlin.de}
\author*[1]{Daniel Schick}\email{schick@mbi-berlin.de}

\affil[1]{Max-Born-Institut für Nichtlineare Optik \& Kurzzeitspektroskopie, Max-Born-Str. 2A, 12489 Berlin, Germany}
\affil[2]{Helmholtz-Zentrum Berlin für Materialien \& Energie GmbH, Hahn-Meitner-Platz 1, 14109 Berlin, Germany}
\affil[3]{Dipartimento di Ingegneria Elettrica, Università di Napoli "Federico~II", Via Claudio 21, 80125 Napoli, Italy} 
\affil[4]{Laboratory of Nanoscale Magnetic Materials \& Magnonics, Institute of Materials (IMX), École Polytechnique Fédérale de Lausanne (EPFL), 1015 Lausanne, Switzerland}
\affil[5]{Institute of Electrical \& Micro Engineering (IEM), École Polytechnique Fédérale de Lausanne (EPFL), 1015 Lausanne, Switzerland}
\affil[6]{Institut für Physik \& Astronomie, Technische Universität Berlin, Straße des 17. Juni 135, 10623 Berlin, Germany}

\begin{acronym}
\acro{GC}{grating coupler}
\acro{CPW}{coplanar waveguide}
\acro{BLS}{Brillouin light scattering}
\acro{RIXS}{resonant inelastic X-ray scattering}
\acro{STXM}{scanning transmission X-ray microscopy}
\acro{SAXS}{small-angle X-ray scattering}
\acro{MMM}{magnon momentum microscopy}
\acro{SEM}{scanning electron microscopy}
\acro{YIG}{yttrium iron garnet}
\acro{GGG}{gadolinium gallium garnet}
\acro{CCD}{charge-coupled device}
\acro{DE}{Damon-Eshbach}
\acro{Py}{permalloy}
\acro{BWV}{backward-volume}
\acro{4MS}{four-magnon scattering}
\acro{XMCD}{X-ray magnetic circular dichroism}
\acro{XMLD}{X-ray magnetic linear dichroism}
\acro{PSSW}{perpendicular standing spin wave}
\acro{SW}{spin wave}
\acro{RF}{radio frequency}
\acro{DC}{direct current}
\acro{SWM}{spin-wave model}
\acro{SNR}{signal-to-noise ratio}
\acro{FIB}{focused ion beam}
\acro{NV}{nitrogen-vacancy}
\acro{STEM}{scanning transmission electron microscopy}
\acro{EELS}{electron energy loss spectroscopy}
\end{acronym}

\date{\today}

\abstract{%
Magnons are quantised collective excitations of long-range ordered spins.
At nanometre wavelengths, exchange interactions increasingly govern their dynamics, giving rise to a largely unexplored regime of couplings between magnons and other quasiparticles.
Yet, detecting such short-wavelength spin waves has remained a key experimental challenge.
Here, we introduce \ac{MMM}---a quasi-elastic, resonant magnetic soft-X-ray scattering technique that directly images magnon populations across two-dimensional momentum space.
Owing to its remarkable sensitivity, \ac{MMM} can capture nonlinear magnon--magnon interactions over large regions of the dispersion plane.
Applying \ac{MMM} to the prototypical magnonic material \ac{YIG}, we uncover a rich variety of previously unobserved nonlinear magnon interactions.
With its element specificity, bulk sensitivity, as well as intrinsic access to nanometre-scale wavelengths without frequency limitation, soft-X-ray \ac{MMM} establishes a powerful and versatile platform for exploring short-wavelength and nonlinear magnonics.
}


\maketitle

\acresetall

\section{Introduction}

Spin waves and their quasi-particles, magnons, provide a versatile platform
for exploring alternative concepts for wave-based information processing beyond conventional CMOS technology~\cite{Chumak2015}.
In particular, nonlinear magnonics has emerged as a promising domain for realizing computing schemes that exploit the intrinsic nonlinearity of magnon interactions~\cite{Wang2020-2,Papp2021,Koerber2023}. 
As the magnon wavelength decreases, short-range exchange interactions begin to dominate. 
In the prototypical magnonic material \ac{YIG}, this crossover typically occurs for wavelengths below \qty{100}{nm}.
A major challenge in accessing this largely unexplored regime is the ability to reliably excite and detect such short-wavelength modes~\cite{Flebus2024}.

The excitation of magnons in the sub-100-nm range has been recently demonstrated by the exploitation of spin-torque architectures~\cite{Demidov2010,Demidov2012} and spin textures~\cite{Albisetti2020,Wintz2016} as spin-wave emitters.
Alternatively, magnonic \acp{GC} and ferromagnetic \acp{CPW} can reach the sub-100-nm-wavelength regime by direct electrical microwave excitation of spin waves~\cite{Liu2018,Che2020,Baumgaertl2020,Baumgaertl2023}.

On the detection side, accessing such high magnon frequencies or wave vectors remains challenging and is a relevant and topical area of research. 
Electronic techniques, commonly based on microwave absorption~\cite{Farle1998}, spin-Hall effect~\cite{Cornelissen2015},  spin-wave-transmission~\cite{Ciubotaru2016}, or magnetoresistive detection~\cite{Rossi2025} have emerged as reliable tools providing information on the frequency-dependent magnon amplitude in the GHz regime. 
Light, however, can be used as a more versatile probe of magnons. 
Unlike electronic detection, optical methods do not rely on patterned antennas, contacts, or Hall structures, allowing flexible scattering and spectroscopy geometries under largely unconstrained sample conditions. 
Generally, the interaction of photons with magnons can be described quantum-mechanically by the creation/annihilation of magnons in a Stokes/Anti-Stokes process. 
Due to the conservation of energy and momentum, the scattered photons experience a frequency and momentum transfer from the \ac{SW} of $ f_\mathrm{f} = f_\mathrm{i} \mp f_{\mathrm{SW}}$ and $\vec{k}_\mathrm{f} = \vec{k}_\mathrm{i} \mp \vec{k}_{\mathrm{SW}}$, respectively, with frequency $f$ and wave vector $\vec{k}$ and indices denoting the photon's initial (i) and final (f) states.
\ac{BLS} in the optical range has been a very successful tool exploiting this inelastic scattering approach in frequency space for magnons up to the few-GHz range with momentum and spatial resolution~\cite{Demokritov2001, sebastian2015a, Krcma2025}. %

To truly access the sub-100-nm regime, probing by short-wavelength X-rays is essential.
Pioneering techniques such as \ac{RIXS}~\cite{Hill2008, Suzuki2019} and \ac{STXM}~\cite{Wintz2016} have enabled significant insights, the former in frequency and momentum space, the latter in real space and time domain.
However, the direct detection of nonlinear interactions of sub-100-nm magnons across the entire dispersion plane remains a critical challenge. 
Existing methods, while powerful within their specific domains, face fundamental constraints in sensitivity, efficiency, and accessible phase space, leaving a substantial \emph{detection gap} in magnonics where key dynamic processes unfold unseen. 
To bridge this gap, we here present the development of \acf{MMM}, an advanced X-ray technique to directly \emph{image} magnon populations in two-dimensional momentum space. 
We demonstrate the unique capabilities of this approach by our experimental observation of nonlinear magnon-magnon scattering in the exchange-dominated regime, opening a new window into the exploration of magnon dynamics beyond previous limitations.

\section{Magnon Momentum Microscopy}

\begin{figure*}
  \centering
  \includegraphics[width=0.98\textwidth]{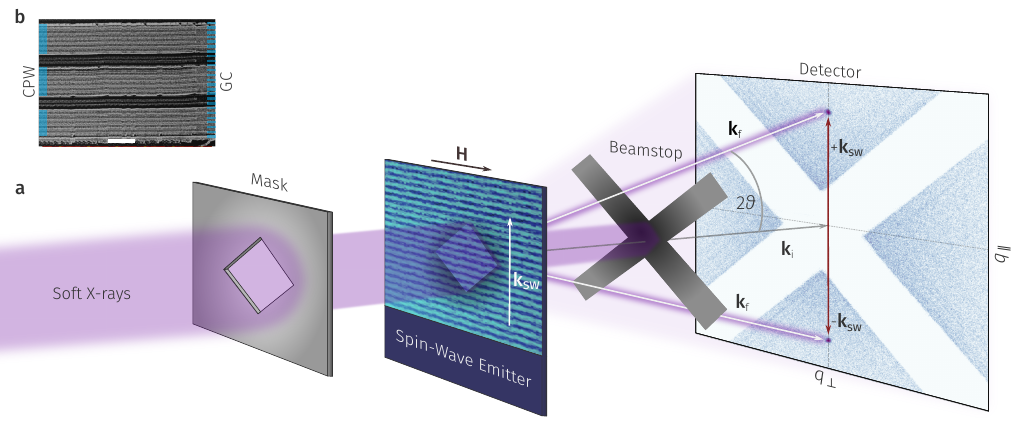} %
  \caption[]{\textbf{Soft-X-ray magnon momentum microscopy.}
  \textbf{a}, Schematics of the soft-X-ray scattering geometry.
  Plane-wave magnons in \acl{DE} configuration ($\mu_0 \vec{H} \perp \vec{k}_\mathrm{SW}$) propagate away from the spin-wave emitter, as indicated by the bluish, out-of-plane magnetisation contrast.
  Quasi-elastic, resonant magnetic scattering with the magnons results in $+1$\textsuperscript{st} and $-1$\textsuperscript{st}-order diffraction peaks on the detector, revealing the magnon wave vector, $\vec{k}_\mathrm{SW}$, directly in momentum space.
  \textbf{b}, \Ac{SEM} image of the spin-wave emitter, consisting of a \acf{CPW} and thin permalloy stripes, forming the \acf{GC}, as indicated by the blue areas.
  Scalebar, \qty{2}{\micro\meter}.
  }
  \label{fig:fig1_setup_nutshell}
\end{figure*}

In this work, we study magnons down to the sub-100-nm regime by directly accessing their wave vectors. 
We do so by employing quasi-elastic resonant magnetic soft-X-ray scattering as a highly sensitive and effective tool to probe magnons in momentum space. 
In analogy to recent works on X-ray scattering with temporally coherent phonons~\cite{boja2013a, Trigo2013, Maznev2021, capotondi2025}, we neglect the small energy transfer between the magnons (sub-eV) and soft-X-ray photons (hundreds of eV), which in a classical picture stems from a Doppler shift of the moving spin wave. 
Accordingly, we can describe the spin waves as a quasi-static periodic magnetic modulation during the interaction with a soft-X-ray photon~\cite{miedaner2024}. 
This modulation effectively forms an absorption grating when tuning the photon energy to electronic resonances exhibiting an \ac{XMCD}~\cite{Kortright2013}.
This concept of \acf{MMM} is sketched in Fig.~\ref{fig:fig1_setup_nutshell}a for probing propagating spin waves via soft-X-ray scattering. 
The requirements for this soft-X-ray experiment are modest: 
A moderate beam focus of tens of micrometres is sufficient. 
Crucially, the photon energy must match a magnetically sensitive absorption edge to achieve high scattering contrast. 
Although this absorption contrast relies on a \emph{circular} dichroism, the magnetic \emph{scattering} does not require a defined X-ray polarisation~\cite{Kortright2013, Lunin2025}.
In a simple transmission scheme, a beamstop prevents the direct beam from hitting the two-dimensional soft-X-ray detector, and a proximity mask defines the probed area of interest and avoids any topographic scattering, resulting in negligible background scattering. 
For spin waves with a well-defined wave vector $\vec{k}_\mathrm{SW}$ (with wavelength $\lambda_{\mathrm{SW}} =  2\pi/|\vec{k}_{\mathrm{SW}}|$), magnetic diffraction peaks of $+1$\textsuperscript{st} and $-1$\textsuperscript{st} order emerge in the scattering pattern at $\vec{q} = \pm \vec{k}_\mathrm{SW}$, with $\vec{q}$ being the scattering vector. 
This simple scattering relation forms the basis of our momentum microscopy image.

In order to demonstrate and benchmark the capabilities of \ac{MMM}, we excite magnons down to sub-\qty{100}{nm} by coupling a microwave-generated magnetic field (with \acl{RF} $f_\mathrm{RF}$) from a \acf{CPW} to an underlying \acf{GC} structure, integrated onto a 100-nm-thick \ac{YIG} film on a \ac{GGG} substrate~\cite{Baumgaertl2020, Baumgaertl2023}, as pictured in Fig.~\ref{fig:fig1_setup_nutshell}b.
Some of us recently demonstrated the feasibility of this approach for the generation of propagating sub-100-nm magnons at frequencies up to $\approx \qty{8}{GHz}$ by \ac{STXM}~\cite{Baumgaertl2020}, %
using a sister sample of the one investigated here. 
The spin-wave excitation is entirely performed in the \ac{DE} configuration, i.e., the magnetic field is applied parallel to the \ac{CPW} and \ac{GC} structures, see Fig.~\ref{fig:fig1_setup_nutshell}a. 
In consequence, the propagation direction of the excited magnons is expected to be purely perpendicular to the field, i.e., $\vec{k}_\mathrm{SW} \perp \mu_0 \vec{H} $, and the momentum transfer to the soft X-rays occurs along the $q_{\parallel}$ direction.
More details on the experimental setup and sample are given in the Methods section. %

The quasi-background-free detection of magnetic scattering with the \ac{MMM} technique enables unparalleled sensitivity  up to high magnon wave vectors.
We observe a distinguishable diffraction signal down to an excitation power of at least \qty{-34}{dBm} at an integration time of $t_{\mathrm{int}} = \qty{30}{s}$ (see Extended Data Fig.~\ref{fig:sensitivity}). %
For comparison, more than a thousand times more microwave power (\qty{-3}{dBm}) was required to properly distinguish the spin-wave signal from background noise in STXM~\cite{Baumgaertl2020} from a sister sample at reasonable integration times.

Direct access to the intensity and wave vector of magnons down to sub-100-nm wavelength within only seconds of integration time is the first main achievement of this work. 
In the following, we demonstrate that the strength of our \ac{MMM} concept stretches beyond simple diffraction from well-defined magnons, revealing novel nonlinear processes uncovered by the two-dimensional accessibility of momentum space. %

\section{Nonlinear Magnon Processes in two-dimensional Momentum Space}

\begin{figure}
  \centering
  \includegraphics[width=0.96\columnwidth]{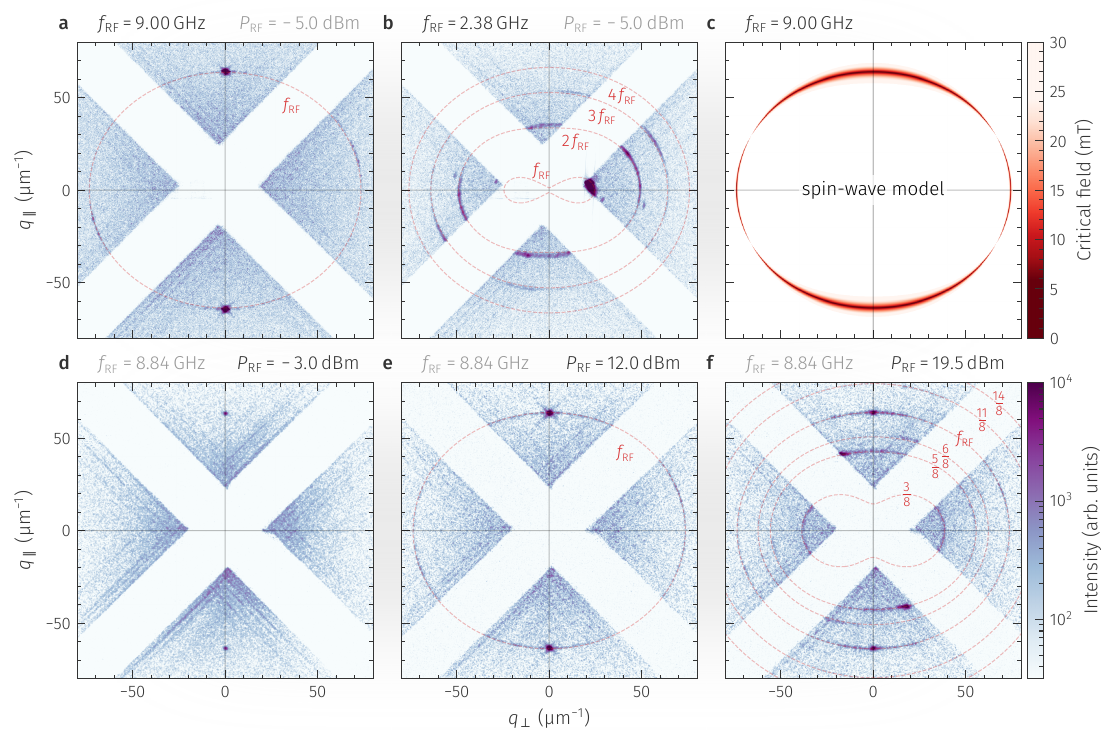}%
  \caption[]{\textbf{Imaging nonlinear magnon processes in momentum space.}
  \textbf{a}, Elliptical dispersion ring resulting from nonlinear magnon--magnon scattering of the directly excited \acl{DE} spin waves into modes of arbitrary propagation direction at $f_\mathrm{RF}=\qty{9.00}{GHz}$.
  \textbf{b}, Higher harmonics and mode redistribution at $f_\mathrm{RF}=\qty{2.38}{GHz}$; red dashed lines indicate theoretical dispersion curves for $f_\mathrm{SW} = n\, f_\mathrm{RF}$.
  \textbf{c}, Calculated \acl{RF} critical field for nonlinear scattering from the \acl{SWM}, evaluated for the excitation power corresponding to \qty{-5.0}{dBm} used in \textbf{a} and \textbf{b}.
  \textbf{d--f}, Power-dependent transition at $f_\mathrm{RF}=\qty{8.84}{GHz}$ (bias field $\mu_0 H=\qty{20}{mT}$ deviating from other shown measurements) from linear excitation (\textbf{d}) to a nonlinearly excited elliptical ring (\textbf{e}) and higher/fractional harmonics as indicated (\textbf{f}).   
  Red dashed lines represent theoretical dispersion iso-frequency curves for the most prominent harmonics. 
  Panels \textbf{a, b, d--f} share the intensity scale.  
  } 
  \label{fig:nonlinear_exp_theo}
\end{figure}

The excitation efficiency of magnons by the electronically-driven \ac{CPW} and \ac{GC} structure strongly depends on the driving frequency through the corresponding geometrically favoured wave vectors~\cite{Baumgaertl2020} %
(excitation spectrum is found in Extended Data Fig.~\ref{fig:CPW_efficiency}a). %
Hence, at frequencies where the excitation efficiency is high, elevated magnon populations are generated, which foster pronounced nonlinear magnon interactions~\cite{Suhl1957}.
In Fig.~\ref{fig:nonlinear_exp_theo}a and b, we show the \ac{MMM} images at the exact same magnetic field and excitation power as shown in Fig.~\ref{fig:fig1_setup_nutshell}a ($f_\mathrm{RF}=\qty{8.68}{GHz}$), but for two distinct excitation frequencies. 
We provide a movie of the dataset containing the entire frequency sweep as Supplementary File. 
At high frequencies around $f_\mathrm{RF}=\qty{9.00}{GHz}$, mainly involving a higher \ac{GC} mode, the two anticipated diffraction peaks in the $q_{\parallel}$ direction are accompanied by a distinct elliptical scattering ring (Fig.~\ref{fig:nonlinear_exp_theo}a). 
This ring clearly indicates the population of magnon modes across all directions of propagation.
Remarkably, this effect comprises that the directly excited \ac{DE} mode along $q_{\parallel} \approx \qty{64}{\per\micro\meter}$  ($\lambda_\mathrm{SW} \approx \qty{98}{nm}$)  is also scattered into \ac{BWV} magnon modes of larger wave vector along $q_{\perp} \approx \qty{73}{\per\micro\meter}$ ($\lambda_\mathrm{SW} \approx \qty{86}{nm}$). 

To elucidate the underlying nonlinear magnon dynamics, we develop the \acf{SWM}~\cite{Suhl1957}, detailed in Supplementary Information. 
Our analysis reveals that the mechanism responsible for populating magnon modes across all directions is a spin-wave parametric resonance driven by four-magnon scattering.
In this process, two directly excited \ac{DE} modes couple to two secondary magnons of arbitrary wave vector $\pm\vec{k}{_\mathrm{SW}}$. 
The coupling strength is proportional to the population of the \ac{DE} modes. 
Parametric excitation becomes possible only when the population of the \ac{DE} modes reaches a threshold value corresponding to a threshold \ac{RF} field (or power) amplitude.
The critical field for the parametric excitation--threshold field distribution in momentum space can be derived from the SWM (see Supplementary Information) and is presented in Fig.~\ref{fig:nonlinear_exp_theo}c for an excitation frequency of $f_\mathrm{RF} = \qty{9.00}{GHz}$.
Importantly, the observed parametric instability mechanism is fundamentally distinct from the commonly invoked Suhl instability~\cite{Suhl1957}, in which the uniform mode ($k_\mathrm{SW}=0$) parametrically excites spin-wave modes with $k_\mathrm{SW}\neq0$.
To our knowledge, this represents the first direct observation of an omnidirectional magnon population arising from such a four-magnon parametric process. %
The calculations of the critical \ac{RF} field amplitude for parametric excitation reveal that, for the \ac{RF} power values used in the experiments, only  spin-wave modes satisfying the resonance condition $f_\mathrm{SW} = f_\mathrm{RF}$ can be parametrically excited, implying that all magnons on the elliptical diffraction ring in Fig.~\ref{fig:nonlinear_exp_theo}a oscillate at the same frequency.
Furthermore, the magnetization deflection angle at the parametric-instability threshold remains small ($\theta_\mathrm{max}\approx \ang{2}$ at $f_\mathrm{RF} = \qty{2.38}{GHz}$ and $\theta_\mathrm{max}\approx \ang{10}$ at $f_\mathrm{RF} = \qty{9.00}{GHz}$; see Supplementary Information). 
\ac{MMM} can access the entire magnon dispersion plane in a single acquired image, and we find near-perfect match between the theoretical direction-dependent spin-wave dispersion (red dashed line) and the scattering ellipse of the nonlinearly formed modes. 
Details on the theoretical calculations are provided in the Methods section.

We now turn the attention to lower excitation frequencies ($f_\mathrm{RF} = \qty{2.38}{GHz}$), where the excitation efficiency is even larger due to the direct coupling of the \ac{CPW}-generated field to the spin waves (without involving the \ac{GC} conversion). 
The corresponding \ac{MMM} image is shown in Fig.~\ref{fig:nonlinear_exp_theo}b and reveals the nonlinear generation of higher-order modes of similar elliptical shape as in panel a. 
These additional modes correspond to higher harmonics of the fundamental mode $f_\mathrm{SW} = n\,f_\mathrm{RF}$ ($n=1, 2, 3, 4)$ as indicated by the comparison to the theoretical dispersion rings (red dashed lines). 
In this low-frequency, long-wavelength regime, the intense fundamental (first order) almost entirely disappears behind the beamstop due to the optimization of the detection geometry for larger momentum transfer in our experiment.  
However, we observe that the directional dependence of the second order slightly deviates from the calculated dispersion curve. 
We attribute this to the high population of these modes and, consequently, the influence of a nonlinear frequency shift~\cite{Suhl1957}. 
Accordingly, the deviation relaxes for the less intense third and fourth orders, which again show a good match to theory.

We further investigate the threshold behaviour of the nonlinear magnon interactions and acquire \ac{MMM} images of an extended series of excitation powers. 
While we again provide the image series as a movie as Supplementary File, Fig.~\ref{fig:nonlinear_exp_theo}d--f picture selected snapshots of a clear hierarchy of nonlinear effects in the exchange-dominated magnon regime. 
At lower power, single diffraction peaks are detected, reflecting the linear propagation of the excited spin wave. 
As the power increases, a fundamental dispersion ring forms, followed by additional parametric processes that generate spin waves at fractional harmonics of the fundamental frequency ($f_\mathrm{SW}=\frac{m}{8}f_\mathrm{RF}$,
with $m=3$ to $14$, except for $m=4$ and $10$).  
The appearance of these modes highlights the deeply nonlinear character of the strongly driven system. 
While a detailed theoretical understanding of these fractional harmonics is beyond the scope of this study, they underscore the rich information content accessible with MMM and remain a topic for future investigation.

\section{High-Resolution Mapping of the Spin-Wave Dispersion }

\begin{figure}
  \centering
  \includegraphics[width=0.96\columnwidth]{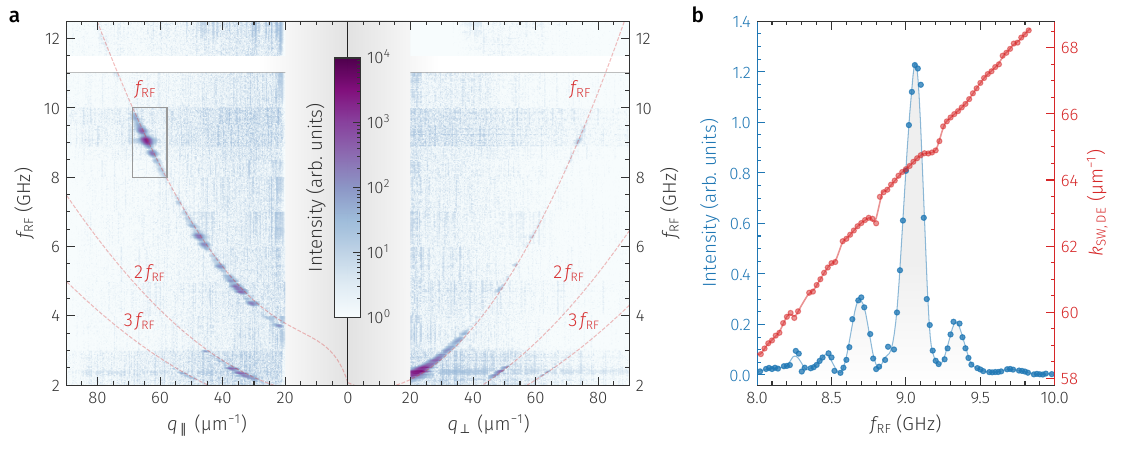} %
  \caption[]{\textbf{Extracted magnon dispersions.}
  \textbf{a}, Dispersion along $q_{\parallel}$ and $q_{\perp}$, corresponding to the \acs{DE} and the \acs{BWV} spin-wave modes, respectively. 
  The \acs{BWV} mode arises from nonlinear scattering. 
  Red dashed lines indicate theoretical dispersions for $f_{\mathrm{SW}} = n\,f_{\mathrm{RF}}$ with $n=1, 2, 3$.
  \textbf{b}, Zoom into the \ac{DE} dispersion at high frequencies (region marked in \textbf{a}), showing intensity maxima and discrete wave vector \emph{jumps} due to the grating coupler geometry. Lines are guides to the eye.
  } 
  \label{fig:dispersion}
\end{figure}

Determining the spin-wave dispersion relation at sub-\textmu m wavelengths is an experimental challenge, as traditional methods typically fail to capture the magnonic behaviour in this regime. 
In contrast, MMM directly accesses reciprocal space, naturally enabling the dispersion to be resolved with an unprecedented combination of precision and range. 
In Fig.~\ref{fig:dispersion}a, we extract the experimental dispersion curve directly from the extended frequency series, with representative snapshots presented in Figs.~\ref{fig:fig1_setup_nutshell}a and \ref{fig:nonlinear_exp_theo}a,b. 

The dispersion for a given propagation direction is readily obtained from directional lineouts of the diffraction data. 
As the magnon generation is not limited to the primarily excited \ac{DE} mode, the nonlinear formation of the fundamental dispersion ring allows access to arbitrary momentum directions. 
We show the dispersion of the prominent \ac{DE} mode along $q_{\parallel}$ (left) and of the \ac{BWV} mode along the $q_{\perp}$ (right) in Fig.~\ref{fig:dispersion}a. 
The data cover a broad range of the dispersion, clearly revealing its general parabolic form ($f_\mathrm{SW}\propto k_{\mathrm{SW}}^2$), a hallmark of the exchange-dominated nature of the magnons. 
The theoretical dispersion curves (red dashed lines) show a very high agreement with the experimental data for the fundamental as well as the second and third harmonics.
Note that the higher-harmonic branches appear at lower frequencies than the fundamental because the dispersion in Fig.~\ref{fig:dispersion}a is shown as a function of the excitation frequency $f_\mathrm{RF}$.
The actual magnon frequencies, however, obey $f_\mathrm{SW} = n\,f_\mathrm{RF}$, and \ac{MMM} measures the corresponding wave vectors.

As noted earlier, the \acp{CPW} and \acp{GC} excitation scheme preferentially generates magnons with specific wave vectors, defined by the grating geometry 
\cite{Yu2013,Baumgaertl2020}.
The resulting scattering intensities, which correspond to the spin-wave intensity, directly reflect this wave-vector-dependent efficiency. 
As a result, the \ac{DE} dispersion curve shows pronounced intensity variations, with several principal intensity peaks, corresponding to favoured wave vectors, such as the fundamental $k_1$-mode of the \ac{CPW} at \qty{2.4}{GHz} and the 2\textsuperscript{nd}-order mode of the \ac{GC} ($2G$) at \qty{4.7}{GHz}~\cite{Baumgaertl2020}. 
We provide an extended frequency-dependent lineout as Extended Data Fig.~\ref{fig:full_spectrum}.

In Fig.~\ref{fig:dispersion}b, we focus on the frequency band related to the efficient excitation of large wave vectors mediated through the \ac{GC} periodicity $a = \qty{400}{nm}$.  %
The dominant diffraction peak corresponds to $k_\mathrm{SW,DE}= 4G = 8\pi/a$, with weaker peaks arising from modulations of $4G$ with other \ac{CPW}-preferred wave vectors. 
The extracted wave vectors (red curve) show a discontinuous frequency dependence, where spin waves are predominantly excited at discrete wave vectors, reflecting confinement by the excitation geometry. 
Transitions between these preferred values are marked by abrupt jumps and, hence, slight local bending of the dispersion, indicating pulling effects toward geometrically favoured modes. 
This level of detail, observed in a planar film region micrometres away from the emitter, highlights the sensitivity and momentum resolution of \ac{MMM}.

\section{Discussion}

A central motivation for the development of \ac{MMM} is to enable the direct detection of magnons with nanometre-scale wavelengths entering the exchange-dominated regime---a domain largely inaccessible with existing techniques.
In this first demonstration of \ac{MMM}, we directly excite and detect magnon wave vectors up to $q_{\parallel} = \qty{79.2}{\per\micro\meter}$ ($f_\mathrm{RF} = \qty{12.1}{GHz}$, $\lambda_\mathrm{SW} \approx \qty{79}{nm}$), and observe even higher wave vectors for nonlinearly excited magnon modes reaching $q_{\parallel} = \qty{93.8}{\per\micro\meter}$ ($\lambda_\mathrm{SW} \approx \qty{67}{nm}$), as shown in Fig.~\ref{fig:nonlinear_exp_theo}f. 
The upper limit for accessible wave vectors in \ac{MMM} is ultimately set by the overall diffraction limit, specifically, by the wavelength of the scattered soft-X-ray photons, $\lambda_\mathrm{ph} \approx \qty{1.75}{nm}$, corresponding to $2k_\mathrm{ph} \approx \qty{7200}{\per\micro\meter}$.
In our current setup, the maximum detectable wave vector is further limited by the detector geometry, while the beamstop obscures features at $q < \qty{20}{\per\micro\meter}$. 
However, these setup parameters are readily adaptable to specific experimental requirements.

Other detection techniques are also pushing into the sub-100-nm regime, but each comes with trade-offs in terms of resolution, sensitivity, or data interpretation. 
Near-field optical approaches, for example, have been proposed to overcome the diffraction limit of conventional magneto-optical Kerr and \ac{BLS} techniques by using evanescent waves or sharp tips to access the sub-wavelength regime \cite{yao2020a,Krcma2025}. 
However, the near-field signal is often highly complex and requires extensive electromagnetic simulations for interpretation. 
\ac{STXM} has recently demonstrated access to magnons with wavelengths near \qty{100}{nm}~\cite{Wintz2016,Baumgaertl2020}. 
While powerful in resolving real-space spin dynamics, the sensitivity of \ac{STXM} to short (and fast) spin-waves is inherently limited by its spatial and temporal resolution, which is dictated by the X-ray optics and pulse duration, respectively. 
\ac{RIXS} offers access to large parts of the Brillouin zone, probing magnon dispersions up to THz frequencies and \r{A}ngström wavelength \cite{Hill2008, Suzuki2019}. 
However, as an energy-domain method, it is limited in detecting low-energy magnons (\qty{<100}{GHz}) requiring  sub-meV resolution. In addition, \ac{RIXS} measurements are extremely photon-hungry and slow due to inherently weak inelastic scattering cross-sections and losses of dispersive optics.

Magnon probes other than light are often susceptible to variations in the sample environment or rely on stringent external conditions. 
For example, the resonance frequency of \ac{NV} centres shifts in external magnetic fields~\cite{bertelli2020, zhou2021}, complicating measurements. 
Likewise, the impressive progress achieved by combining \ac{STEM} and \ac{EELS} to access antiferromagnetic magnons across the full Brillouin zone with meV energy resolution~\cite{kepaptsoglou2025} relies on strong magnetic electron lenses, which are difficult to reconcile with most ferromagnetic spin-wave materials.

What sets MMM apart from other approaches is its combination of high sensitivity, rapid acquisition, and direct 2D access to momentum space without frequency limitations---all within a relatively simple experimental setting. 
These capabilities originate from the underlying scattering process, where spin-wave-induced magnetic modulations form an effective absorption grating. 
While inelastic processes may contribute~\cite{Madami2011}, their cross-sections are orders of magnitude weaker than the elastic forward scattering. 
Operating directly in reciprocal space, \ac{MMM} bypasses the limitations of inefficient X-ray imaging or dispersion optics, with 2D diffraction patterns that directly encode magnon wave vectors and intensities, eliminating the need for complex reconstruction or modelling.

The efficiency, sensitivity, and direct momentum-space access of MMM enable the study of spin-wave phenomena far beyond the reach of existing techniques, most notably nonlinear magnon interactions at nanometre wavelengths. 
Remarkably, we observe such nonlinear effects in \ac{YIG}, a textbook model system in magnonics. 
Specifically, we directly detect a four-magnon scattering process that parametrically redistributes the \ac{DE} mode into an omnidirectional population of secondary magnons, forming a characteristic elliptical ring in momentum space. 
To interpret this behaviour, we developed a spin-wave model that reformulates the classical Landau–Lifshitz description in terms of spin-wave amplitudes, enabling a clear analysis of the nonlinear interaction terms.
The model captures a parametric instability mechanism distinct from conventional Suhl processes and predicts instability thresholds consistent with our observations. 
That such rich and previously unresolved physics emerges in a prototypical material like YIG highlights the unique capability of MMM to reveal hidden aspects of magnon dynamics. 

\section{Perspective}

\Acl{MMM} offers a powerful new approach for exploring nonlinear spin-wave interactions, mode coupling, and wave-vector-resolved scattering processes across a broad range of magnonic systems and geometries. 
Its high sensitivity, element specificity, and direct momentum-space access make it well-suited for studying multilayer structures, buried interfaces, and engineered magnonic devices. 
The technique is compatible with a wide variety of magnon excitation schemes and sample environments.
\Ac{MMM} can readily be adapted towards material systems with shorter spin-wave propagation lengths than \ac{YIG} by reducing the soft-X-ray-magnon interaction volume using smaller X-ray apertures or dedicated focussing schemes down to the sub-\textmu m regime. 
It can be extended to detect magnons up to THz frequencies~\cite{Hortensius2021,schonfeld2025} in antiferromagnetic materials by accessing a magnon-induced dynamic net magnetization either by \ac{XMCD}~\cite{blank2023}, or depending on the geometry by exploiting \ac{XMLD} as a contrast mechanism.
Time-resolved implementations using short-pulsed X-ray sources, particularly including laser-based laboratory sources, can provide direct insight into frequency-resolved magnon phenomena on ultrafast timescales~\cite{Trigo2013, miedaner2024, capotondi2025}. By enabling direct access to magnon distributions in momentum space, MMM establishes a new experimental paradigm for investigating nonlinear spin-wave dynamics beyond the reach of conventional techniques.

\newpage

\section{Methods}
\acresetall
\subsection{YIG Sample}

The \ac{YIG} 100-nm-thick layer is grown on a \ac{GGG} substrate via liquid-phase epitaxy by Matesy GmbH, Jena, Germany. 
Onto the YIG film, a $20\,$nm Py (Ni$_{81}$Fe$_{19}$) film is deposited via e-beam evaporation, which is subsequently patterned via e-beam lithography to define the \acp{GC} in a negative hydrogen silsesquioxane resist. 
The \ac{Py} stripes of the \ac{GC} are chosen to have a width of \qty{200}{nm} and a periodicity of $a = \qty{400}{nm}$. 
The defined \ac{GC} structures are perpetuated by Ar$^+$ ion beam etching, optimised for minimising the overetching into the \ac{YIG} film. 
The \acp{CPW} are fabricated above the \ac{GC} by using positive PMMA/MMA e-beam resist and evaporation of Ti(5)/Cu(110) onto the patterned structure (nanometre thickness in brackets). 
A \ac{SEM} image of the fabricated structures is shown in Fig.~\ref{fig:fig1_setup_nutshell}b. %

X-ray transparency of the sample is achieved by thinning the \ac{GGG} substrate. 
This process involves initially mechanically thinning the substrate from its backside, followed by local refinement using focused Ga$^+$ ion beam etching~\cite{Foerster2019}. 
To restrict the region probed by the X-ray beam, we used a square-shaped mask that defines the illuminated area on the sample. 
Its main purpose is to confine the illumination to a topographically uniform region, thereby suppressing unwanted topographic (charge) scattering from surrounding structures such as the \ac{CPW} or \ac{GC}, which would otherwise obscure the magnetic scattering from the spin waves. 
The mask was fabricated from a Cr/Au multilayer (total thickness $\approx \qty{1}{\micro m}$) deposited on a SiN membrane, into which a $\qtyproduct{16 x 16}{\micro\meter}$  square aperture was milled using a Ga$^+$ \ac{FIB}. 
It was then transferred onto the front side of the sample and attached by \ac{FIB}-assisted Pt deposition. 
The aperture defines the illuminated detection region in the bare \ac{YIG}, with its centre located approximately \qty{11}{\micro\meter} away from the \ac{CPW}. 
The SiN support electrically insulates the metallic mask from the CPW, preventing any shorting.

Introducing such an aperture inevitably produces additional scattering from the mask edges. 
To minimize this effect, the square-shaped mask was designed such that the edge-induced scattering occurs primarily along two perpendicular directions defined by the mask edges. 
These scattering streaks are effectively blocked by a cross-shaped beamstop placed in the detection path. 
For this reason, the mask was intentionally rotated by $\ang{45}$ with respect to the expected \ac{DE} magnon wave-vector direction $q_{\parallel}$. 
This geometry ensures that the regions affected by mask-edge scattering coincide with the beamstop coverage and are thus blanked out in the \ac{MMM} images, while the magnetic scattering signal remains unaffected.
In future implementations of the method, the mask may be substituted by a well-focused and well-defined X-ray beam properly positioned on the sample.

\subsection{Experimental Setup}

We employ resonant magnetic scattering as depicted in Fig.~\ref{fig:fig1_setup_nutshell}a and described in the main text. 
The experiments were conducted with our mobile MAXI endstation at the beamline P04 of the synchrotron-radiation facility PETRA~III at DESY, Hamburg, and at the beamline UE52-SGM of the BESSY~II storage ring at HZB, Berlin. 
The photon energy is tuned to $E_\mathrm{ph}=\qty{708}{eV}$, selecting the maximum \ac{XMCD} contrast at the Fe~L$_3$ ($2p \rightarrow 3d$) absorption edge of the \ac{YIG} film. 
We used circularly polarised X-rays at PETRA III and linear horizontal polarisation at BESSY.  
The propagating magnons are excited by applying a \ac{RF} electrical signal to the \ac{CPW}. 
The microwave power is chosen to be $P_\mathrm{RF}=\qty{-5.0}{dBm}$ if not stated differently. 
An in-plane magnetic bias field of \qty{30}{mT} was applied by an electromagnet, directed along the \ac{CPW} in \ac{DE} configuration. 
As an exception, the data shown in Fig.~\ref{fig:nonlinear_exp_theo}d--f was measured at \qty{20}{mT}.
The relative distances between the sample, beamstop, and CCD X-ray detector define the accessible wave vectors in the experiment. 
Here, we choose the sample-to-CCD distance to $z_0 = \qty{260}{mm}$, which defines the scattering signal position on the CCD chip through $d = k_{\mathrm{SW}} \, z_0/k_i$, with $k_{\mathrm{SW}}$ the wave number of the magnons, $k_i = 2\pi E_\mathrm{ph}/(hc)$ the wave number of the incident soft X-rays. 
Adjusting these parameters and the beamstop position allows further flexibility regarding the accessibility of magnon wave vectors. 

\subsection{Data Analysis}

The raw data, such as shown in Fig.~\ref{fig:fig1_setup_nutshell} and \ref{fig:nonlinear_exp_theo}, are obtained by directly evaluating the CCD images acquired. 
For each dataset, we subtract a background image, with the \ac{RF} excitation switched off, from the image containing magnetic information (\ac{RF} excitation on). 
The background accounts for the detector offset and remaining topographic scattering. 
Prior to subtraction, we apply a sub-pixel drift correction between both detector images.  
To obtain the intensity and momentum-space position of the scattering peaks in the 2D scattering plane, they are fitted using two-dimensional Gaussian profiles (used in Fig.~\ref{fig:dispersion}b and Extended Data Fig.~\ref{fig:full_spectrum}). 
To extract the dispersion from the data (Fig.~\ref{fig:dispersion}a), we perform lineouts along selected directions in reciprocal space, defined by specific azimuthal angles. 
These directions can be chosen to correspond to the \acl{DE} (\ac{DE}, along $q_{\parallel}$) and \acl{BWV} (\acs{BWV}, along $q_{\perp}$) spin-wave modes (such as shown in Fig.~\ref{fig:dispersion}a), or any other arbitrary orientation. 
By plotting the scattering intensity along these lineouts as a function of the scattering vector, $q$, we obtain intensity profiles that reflect the spin-wave characteristics.

\subsection{Nonlinear Spin-Wave Dynamics}

In the \ac{MMM} images, we observe that above a certain \ac{RF} power, $P_\mathrm{RF}$, and for a fixed excitation frequency, $f_\mathrm{RF} = \omega_\mathrm{RF}/(2\pi)$, spin waves begin to propagate in directions different from that of the directly driven mode. 
This behaviour indicates the onset of a spin-wave instability, which can be understood by examining the coupled nonlinear dynamics of magnons in the \ac{YIG} film. 
In the linear regime, magnon modes are uncoupled (see Supplementary Information). 
However, we show that the excitation of spin waves which are not directly driven by the \ac{RF} field arises from a nonlinear mechanism---specifically, a parametric resonance. 
To analyse this, we reformulate the Landau-Lifshitz equation in terms of spin-wave amplitudes, leading to our \ac{SWM}. 
A full derivation of the \ac{SWM}, inspired by the classic work of Suhl~\cite{Suhl1957}, is provided in Supplementary Information. 

Using the \ac{SWM}, we first characterise the system’s nonlinear dynamics and then derive the dispersion relation. 
This analysis reveals that the magnons excited via nonlinear processes, forming the elliptical scattering ring, resonate at the same frequency as the \ac{RF} field. 
Among the nonlinear interaction terms in the \ac{SWM}, we isolate the parametric terms that are resonant with the \ac{RF} field and involve either the \acf{DE} spin waves or the \ac{RF} mode itself. 
The concept is that, prior to the onset of instability, only the \ac{DE} spin waves and the \ac{RF} field exhibit significant amplitudes.

Following an approach similar to Suhl~\cite{Suhl1957}, we derive an expression for the threshold \ac{RF} field required to trigger the parametric spin-wave instability. 
The predicted threshold values align well with those used in our experiments. 
Finally, the distribution of these threshold values across momentum space offers critical insight into the features observed in the \ac{MMM} images, as discussed later in this section.

\paragraph{Dispersion Relation}

In the case of linear and conservative dynamics, spin-wave normal mode amplitudes evolve independently with a frequency given by the following dispersion relation (see Supplemental Information):
\begin{equation}\label{eq:disp_rel}
\omega_k = \gamma \mu_0 M_\mathrm{s}\sqrt{\left(H-N_x + l_\mathrm{ex}^2 k^2 +N_{k,yy}\right)\left(H - N_x + l_\mathrm{ex}^2 k^2 +N_{k,zz}\right)} \, ,
\end{equation}
where $\omega_k$ is the natural frequency in dimensional unit for wave vector $\vec{k} = k_x \hat{\vec{x}}+k_y\hat{\vec{y}}\neq \bm 0$, $\gamma$ is the absolute value of the gyromagnetic ratio, $M_\mathrm{s}$ is the saturation magnetisation, $H = H_\mathrm{DC}/M_\mathrm{s}$ is the uniform, \ac{DC} field applied along the equilibrium direction ($x$), $N_x$ is the demagnetising factor along the $x$ direction, $l_\mathrm{ex} = \sqrt{2A_\mathrm{ex}/\mu_0M_\mathrm{s}^2}$ is the exchange length with $A_\mathrm{ex}$ the exchange stiffness and $\mu_0$ the vacuum permeability, and $N_{k,yy},\,N_{k,zz}$ are the components of the Fourier transform of the demagnetising tensor computed assuming the magnetisation uniform along the thickness of the sample (see Supplementary Information).
In the next section, $\omega_k$ will be indicated as the spin-wave normal modes frequency in dimensionless form (normalised to $\gamma M_\mathrm{s}$).
The material and experimental parameters used for the theoretical modelling---including the calculation of the magnon dispersion relation (Fig.~\ref{fig:nonlinear_exp_theo}a,b,e,f and Fig.~\ref{fig:dispersion}a) and the threshold field for the parametric instability (Fig.~\ref{fig:nonlinear_exp_theo}c)---are listed in Table~\ref{tab:disp_param}.

\begin{table}[h]    
    \caption{Values of the experimental and material parameters for the calculation of the directional dispersion relation and the threshold field of the parametric instability. 
    For the dataset used in Fig.~\ref{fig:nonlinear_exp_theo}a,b,c and Fig.~\ref{fig:dispersion}a, the nominal field value of $\mu_0 H = 30\,$mT is reduced to $27\,$mT for a better curve match in Fig.~\ref{fig:dispersion}a, a value within the precision range of our electromagnet. For the curves in Fig.~\ref{fig:nonlinear_exp_theo}d,e,f, $\mu_0 H = 20\,$mT is used. 
    }\label{tab:disp_param}%
    \begin{tabular}{@{}rccr@{}}
    \toprule
    name & symbol  & unit &  {value}  \\
    \midrule
         gyromagnetic ratio         & $\gamma$          &  \qty{}{\hertz\per\tesla}     & $2\pi \times \num{28.024e9}$  \\
         saturation magnetisation   & $\mu_0 M_\mathrm{s}$ & mT           & 175            \\
         exchange stiffness         & $A_\mathrm{ex}$   &  pJ/m         & 3.65          \\
         film thickness             & $d$               &  nm           & 100           \\
         Gilbert damping            & $\alpha$        &  --           & \num{1e-4}     \\
    \botrule
\end{tabular}
\end{table}

\paragraph{Threshold for Spin-Wave Parametric Excitation}

The \ac{SWM} contains a variety of nonlinear terms that describe magnon–magnon interactions. 
However, not all of these terms are relevant for deriving the threshold field for parametric excitation of spin waves that are not directly driven by the RF field. 
To focus on the relevant dynamics, we derive a simplified model from the \ac{SWM}, retaining only the nonlinear terms associated with four-magnon scattering processes. 
These terms are both parametric and resonant with the RF field and allow us to obtain the following expression for the critical RF field, $H_\mathrm{RF,crit}$, required to parametrically excite a spin wave with wave number $k$:  
\begin{equation}\label{eq:thresh_field_inst_2}
    H_\mathrm{RF,crit}= \mu_0 M_\mathrm{s}\left(\frac{(\omega_\mathrm{RF}-\omega_k)^2+\alpha^2A_k^2}{\left|\sum_{k'\in \{0,k_\mathrm{DE}\}}\frac{\zeta_{kk'}\hat{h}_{k'}^{+2}}{(\omega_\mathrm{RF}-\omega_{k'}-i\alpha A_{k'})^2}\right|^2}\right)^\frac{1}{4}\,.
\end{equation}

Here,  $k_\mathrm{DE}$ is the wave number of the DE spin waves, $\zeta_{kk'}$ is a coefficient describing the coupling between the spin wave with wave number $k$ and the DE spin wave with wave number $k'$, and $\hat{h}_{k'}^{+}$ is the efficiency of the \ac{CPW} and \ac{GC} corresponding to the counterclockwise $k'$-th component of the RF field. 
All the quantities between parentheses are dimensionless, while the critical field is expressed in mT.
In the Supplemental Information, we show that the observed spin-wave parametric instability is connected to the well-known second-order Suhl instability, yet it is distinct and describes a more general case.

The expression of the critical field allows us to derive the threshold field: 
It sets the minimum critical RF field---and therefore RF power---required to initiate the parametric excitation of omnidirectional spin waves. 
For experimental comparison, only this threshold value is relevant. A simple but approximate expression for such a value can be obtained from Eq.\eqref{eq:thresh_field_inst_2} by setting the resonance condition: $\omega_k = \omega_{\tilde{k}}=\omega_\mathrm{RF}$, and $k = \tilde{k}\in\{k_\mathrm{DE}\}$, that produces the following relation:
\begin{equation}\label{eq:thresh_field_inst_3}
    H_\mathrm{RF,thr}=\min_k H_\mathrm{RF,crit}= \mu_0 M_\mathrm{s}\left(\frac{A_{\tilde{k}}}{|B_{\tilde{k}}|}\,\frac{\alpha^{3/2}\omega_\mathrm{RF} \sqrt{A_{\tilde{k}}}}{|\hat{h}_{\tilde{k}}^+|\sqrt{H-N_x}}\right)\,.
\end{equation}
Above the threshold value of the RF field, nonlinear effects cause the excitation of additional spin waves, and the corresponding threshold values for these modes may be altered by the presence of already excited ones.

\paragraph{Experimental Implications}

In materials such as \ac{YIG}, characterized by low magnetic damping ($\alpha \approx 10^{-4}$), only \ac{DE} spin waves with wave number $k'$ satisfying $\omega_{k'} \approx \omega_\mathrm{RF}$ contribute significantly to the denominator in Eq.~\eqref{eq:thresh_field_inst_2}, unless $|\hat{h}_{k'}^{+}| \approx \alpha$.  
From Eq.~\eqref{eq:thresh_field_inst_3}, it follows that the critical \ac{RF} field scales proportionally to $(\alpha \, \omega_\mathrm{RF})^{3/2}$  and inversely with the \ac{CPW} and \ac{GC} excitation efficiency. 
A similar relationship has been reported in Ref.~\cite{Suhl1957}.

An order-of-magnitude analysis shows that when the \ac{CPW} and \ac{GC} efficiency, cf. Extended Data Fig.~\ref{fig:CPW_efficiency}a, is $|\hat{h}_{k'}^{+}|\approx \num{5e-3}$, the threshold field is $H_\mathrm{RF,thr}\approx \qty{1}{mT}$, whereas for maximum efficiency ($\approx \num{2e-2}$), the threshold field decreases to $H_\mathrm{RF,thr}\approx \qty{0.01}{mT}$.
These values are consistent with the experiments: 
the threshold power value at $f_\mathrm{RF} = \qty{9.00}{GHz}$ is $P_\mathrm{RF} \approx \qty{-5.0}{dBm}$ (see Fig.~\ref{fig:nonlinear_exp_theo}a).
This power corresponds, according to the COMSOL calculations presented in the Extended Data Fig.~\ref{fig:CPW_efficiency}b, to an \ac{RF} field amplitude of $\qty{1.3}{mT}$. 
At lower \ac{RF} frequencies, $f_\mathrm{RF} = \qty{2.38}{GHz}$, the same level of \ac{RF} power excites additional higher-order spin-wave modes of frequencies $2f_\mathrm{RF},\,3f_\mathrm{RF}$ and $4f_\mathrm{RF}$ (see Fig.~\ref{fig:nonlinear_exp_theo}b). 
The theoretically estimated threshold field (power) is almost two (one) orders of magnitude smaller than the value at $f_\mathrm{RF}=\qty{9.00}{GHz}$. 
This is due to the higher coupling efficiency of the \ac{CPW} at lower wave vectors. 
Therefore, the excitation of higher-order parametric resonances ($n\,f_\mathrm{RF},\,n=2,3,4$) is consistent with the fact that the \ac{RF} field (power) significantly exceeds the threshold value. 

In the Supplemental Information, we present calculations of the maximum magnetization deflection angle  $\theta_\mathrm{max}$ at the threshold \ac{RF} field for parametric excitation of spin waves. 
For $f_\mathrm{RF} = \qty{2.38}{GHz}$, we find $\theta_\mathrm{max}\approx \ang{2}$, and for $f_\mathrm{RF} = \qty{9.00}{GHz}$, $\theta_\mathrm{max}\approx \ang{10}$.
These results indicate that the growth of the \ac{DE} spin-wave amplitude is self-limited by the onset of parametric excitation. 
The small values of $\theta_\mathrm{max}$ further confirm the consistency of the linear approximation used to describe the \ac{DE} spin-wave dynamics below the threshold.
In Fig.~\ref{fig:nonlinear_exp_theo}a,b, we observe spin waves at frequencies corresponding to integer multiples of the \ac{RF} frequency, confirming the excitation of higher-order parametric resonances beyond the primary one. 
As $\omega_\mathrm{RF}$ deviates from the linear spin-wave dispersion, the critical \ac{RF} field increases sharply, thereby suppressing off-resonant modes. 
This behaviour is consistent with the experimental observation that, for a fixed \ac{RF} power above the threshold, the parametrically excited spin waves occupy a narrow region around the $\omega_k = \omega_\mathrm{RF}$ dispersion curve, with noticeable deviations appearing only at much higher power values.

\backmatter

\section{Acknowledgements}

The experiments were conducted at the beamline P04 of PETRA III operated by DESY (Hamburg, Germany), a member of the Helmholtz Association HGF, and beamline UE52-SGM  of the BESSY II electron storage ring at Helmholtz-Zentrum Berlin, Germany. 
We gratefully acknowledge M.~J.~Huang and M.~Hoesch (DESY) as well as M.~Fondell (BESSY II) for beamline support. 
We also thank the co-authors of Ref.~\cite{Baumgaertl2020} for contributions to earlier work on the sample investigated here.
We thank F.~Büttner for the valuable discussions.
S.W.\ acknowledges financial support from the Helmholtz Young Investigator Group Program (VH-NG-1520). 
S.P.\ acknowledges funding from the HIDA Helmholtz Visiting Researcher Grant No.\ 17804. 
B.P.\ acknowledges funding from the Leibniz Association via Grant No.\ K720/2025 (X-Mag). 
D.S.\ thanks the Leibniz Association for funding through the Leibniz Junior Research Group Grant No.\ J134/2022.

\section*{Declarations}

\begin{itemize}
\item \textbf{Conflict of interest}:
The authors declare no competing interests.
\item \textbf{Data availability}: After acceptance of the manuscript, data and code will be provided by a publicly accessible online repository. 
\item \textbf{Author contribution}:
D.G., B.P., S.W. and D.S.\ conceived the study. 
K.B.\ and M.S.\ prepared the sample. 
S.W., C.K., M.S., J.F., V.D., T.K., and B.P.\ performed the measurements.
S.W., C.K., A.M., and D.S.\ analysed the data. 
S.P.\ developed the spin-wave model and performed the model calculations. 
S.W., S.P., A.M., D.G., B.P., and D.S.\ interpreted the results. 
S.W., S.P., B.P., and D.S.\ drafted the manuscript with input from all other authors. 
Supervision by D.G., S.E., B.P., and D.S. 
\end{itemize}

\newpage 

\begin{appendices}

\section*{Extended Data}

\renewcommand{\figurename}{Extended Data Fig.}
\renewcommand{\thefigure}{E\arabic{figure}}
\setcounter{figure}{0}

\acresetall
\begin{figure}[htp]
  \centering
  \includegraphics[height=5cm]{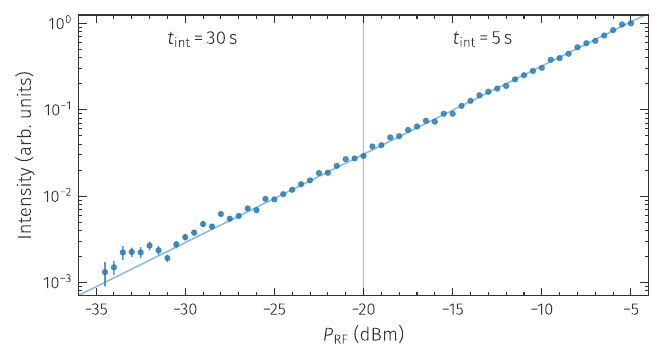}
  \caption[]{Normalised intensity of the $4G$-scattering peak at \qty{9.00}{GHz} for increasing excitation power, $P_\mathrm{RF}$, benchmarking the sensitivity of the \ac{MMM} technique.
  The \acl{SNR} enables detection of the scattering signal down to $P_\mathrm{RF} = \qty{-34}{dBm}$ at reasonable acquisition times of \qty{30}{s} for a single \ac{MMM} image.
  For $P_\mathrm{RF} > \qty{-20}{dBm}$, the integration time is reduced to only \qty{5}{s}.
  The signal follows the expected exponential behaviour (solid blue line). %
  }
  \label{fig:sensitivity}
\end{figure}

\acresetall
\begin{figure}[htp]
  \centering
  \includegraphics[width=\textwidth]{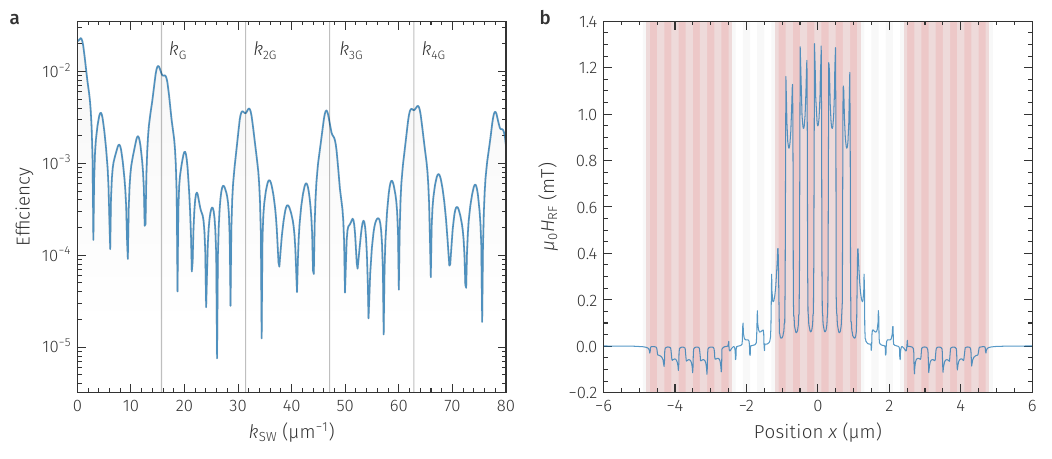}%
  \caption[]{\textbf{Calculated magnon excitation of the \acs{CPW} and \acs{GC}. a} Excitation efficiency of the \acs{CPW} and \acs{GC}.
  The calculation is performed by Fourier transformation of the simulated \acs{RF} in-plane Oersted fields upon microwave excitation. 
  The \acs{CPW} efficiently transfers a wave vector $k_1 = \qty{0.85}{\per\micro\meter}$ to the \acs{YIG} film~\cite{Baumgaertl2020}. 
  The efficiently transferred larger wave vectors mainly result from the \acs{GC} pattern.
  Its natural wave vector is $G = 2\pi/a = \qty{15.7}{\per\micro\meter}$, along with the higher orders $2G$, $3G$, and $4G$.
  \textbf{b} Magnetic excitation field distribution below the \ac{CPW} (red areas) and \ac{GC} (grey areas).
  }
  \label{fig:CPW_efficiency}
\end{figure}

\acresetall
\begin{figure}[htp]
  \centering 
  \includegraphics[width=0.65\textwidth]{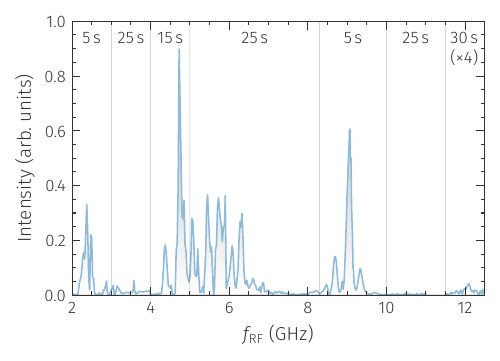}
  \caption[]{\textbf{Excitation spectrum.} 
  Scattering intensities over the entire frequency range $f_\mathrm{RF}$ at an applied field of $\mu_0 H = \qty{30}{mT}$ for the \acs{DE} direction along $q_{\parallel}$ only.
  Hence, in the case of nonlinear mode redistribution and ring formation (see Fig. \ref{fig:nonlinear_exp_theo}), the plotted intensity does not represent the full excitation efficiency. 
  Intensities were normalized to the acquisition time as indicated. 
  For frequencies below \qty{3.30}{GHz}, the data correspond to second-order scattering, while the first order could be directly analysed for higher frequencies. 
  The high-frequency mode around \qty{12.1}{GHz} is additionally multiplied by a factor of four for better visibility.
  Principal peaks at \qty{2.40}{GHz}, \qty{4.70}{GHz}, and \qty{9.10}{GHz} correspond to $k = k_1$, $2G$, and $4G$ modes, respectively.}
  \label{fig:full_spectrum}
\end{figure}

\acresetall
\begin{figure}[htp]
  \centering 
  \includegraphics[width=\textwidth]{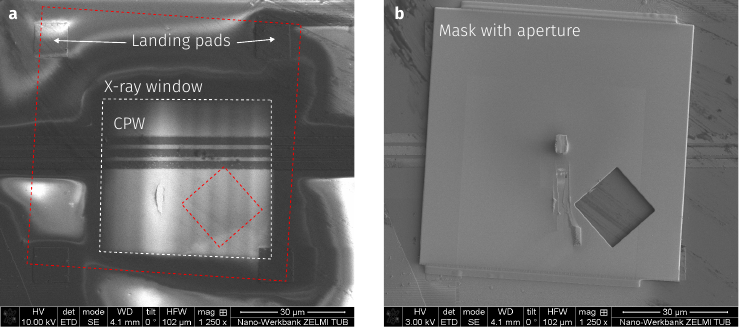}
  \caption[]{\textbf{\acs{SEM} images of the sample layout.} 
  \textbf{a}, Sample area with X-ray-transparent window before attaching the mask.
  The image is recorded with 10-keV electrons to penetrate through the window, which is crossed by the \acs{CPW}. 
  Red dashed lines indicate the later position of the mask and the aperture. 
  Four landing pads were used to slightly elevate the mask above the \acs{CPW}.
  Bright areas at the rim of the image are caused by charging of the insulating \acs{YIG} layer. 
  \textbf{b}, Same sample area as in \textbf{a} after attaching the mask. 
  The image was recorded using 3-keV electrons to highlight surface details.}
  \label{fig:mask}
\end{figure}

\clearpage

\renewcommand{\figurename}{Supplemental Fig.}
\setcounter{figure}{0}
\renewcommand{\thefigure}{S\arabic{figure}}

\section{Supplementary Information}

\subsection{Spin-Waves Model}
In this Supplementary Information, we develop the \ac{SWM} that will be used to explain the observed nonlinear spin-wave dynamics. 
The magnetic sample will be assumed uniformly magnetised along $x$ direction due to the presence of a uniform \ac{DC} magnetic field, simply called \ac{DC}-field in the following, directed in the same direction. 
The spin-wave dynamics is excited by an \ac{RF} field directed always in the plane of the sample but orthogonal to the uniform field, namely the $y$ direction.
The magnetisation evolution is described by the Landau-Lifshitz equation~\cite{Serpico2008}:
\begin{equation}\label{eq:LL}
    \frac{\partial \vec{m}}{\partial t} = -\vec{m}\times \vec{h}_\mathrm{eff} - \alpha\,\vec{m}\times\left(\vec{m}\times \vec{h}_\mathrm{eff}\right)\, ,
\end{equation}
where $\vec{m}(\vec{r},t) = \vec{M}(\vec{r},t)/M_\mathrm{s}(T)$, with $M_\mathrm{s}(T)$ the saturation magnetisation at the temperature $T$, $t$ is the dimensionless time normalized to $(\gamma M_\mathrm{s})^{-1}$, with $\gamma$ the absolute value of the gyromagnetic ratio, $\alpha$ the damping coefficient, and $\vec{h}_\mathrm{eff} = \vec{H}_\mathrm{eff}/M_\mathrm{s}(T)$, with $\vec{H}_\mathrm{eff} = \vec{H}_\mathrm{ex} + \vec{H}_\mathrm{m} + \vec{H} + \vec{H}_\mathrm{RF}$ the effective field given by the sum of the exchange field, magnetostatic field, uniform DC field, and the \ac{RF} field respectively. 
The magnetisation is assumed to be uniform along the film's thickness ($z$ direction). 
This fact permits the substitution of the effective field in the Eq.~\eqref{eq:LL} with that one averaged along the thickness. 
In the following, when we refer to the effective field, its average along the film's thickness is considered.
As a first step, let us derive the \ac{SWM} considering only the conservative magnetisation dynamics.
By using the same formalism of Suhl~\cite{Suhl1957}, we express the unitary magnetisation components in the following way:
\begin{equation}
    \begin{aligned}
        & m_x = \sqrt{1-m_y^2-m_z^2}\,,\\
        & m_y = \sum_k m_{y,k}\, e^{j\vec{k}\cdot \vec{r}}\,,\\
        & m_z = \sum_k m_{z,k}\, e^{j\vec{k}\cdot \vec{r}}\,,
    \end{aligned}
\end{equation}
where $k = (k_x,k_y)$ is a multi-index and $\vec{k}$ is the corresponding two-dimensional vector defined on the $k$-space plane, $(m_{x,k},m_{y,k},m_{z,k})$ are the Fourier transforms of the unitary magnetisation components, and $\vec{r}$ is the in-plane position vector.
Let us introduce the $k$\textsuperscript{th} spin-wave amplitude $a_k = m_{y,k}+i m_{z,k}$ and $a_{-k} = m_{y,k} -i m_{z,k}$. 
In this respect, we can introduce the following auxiliary variables:
\begin{equation}\label{eq:mpm}
\begin{aligned}
   & m_+ = m_y +i m_z = \sum_k a_k\, e^{j\vec{k}\cdot \vec{r}}\,,\\
   & m_- = m_y -i m_z = \sum_k a_{-k}^*\, e^{j\vec{k}\cdot \vec{r}}\,,\\
   & m_x = \sqrt{1-m_+m_-}\approx 1-\frac{1}{2}m_+m_- = 1-\frac{1}{2}\sum_{k,k'} a_{k'}a_{k'-k}^*\, e^{j\vec{k}\cdot \vec{r}}
\end{aligned}  
\end{equation}
where we made the so-called parabolic approximation in the last equation.
The conservative dynamics of the variable $m_+$ is described by the following relation:
\begin{equation}\label{eq:LLmp}
    \frac{\partial m_+}{\partial t} = \sum_k \frac{\mathrm{d}a_k}{\mathrm{d}t}\, e^{j\vec{k}\cdot \vec{r}} = i h_\mathrm{eff,x} m_+ - i m_x(h_\mathrm{eff,y}+ih_\mathrm{eff,z})
\end{equation}
where
\begin{equation}\label{eq:heffp}
\begin{aligned}
    h_\mathrm{eff,y}+ih_\mathrm{eff,z} = & -\frac{(N_y+N_z)}{2}a_0 -\frac{(N_y-N_z)}{2}a_0^*\\
    & + \sum_k e^{-\vec{k}\cdot \vec{r}} \left[h_{\mathrm{RF},k} - \left(l_\mathrm{ex}^2 k^2 + \frac{N_{k,yy}+N_{k,zz}}{2}\right) a_k\right.\\
    & \quad \quad \quad \quad \quad \quad - \frac{N_{k,yy}-N_{k,zz}}{2} a_{-k}^*\\
    & \quad \quad \quad \quad \quad \quad \left.+\frac{N_{k,xy}}{2}\sum_{k'}a_{k'}a_{k'-k}^*\right]\,,
\end{aligned}
\end{equation}
and 
\begin{equation}
\begin{aligned}\label{eq:heffx}
h_\mathrm{eff,x} = \, & H-N_x\left(1-\frac{1}{2}\sum_k\left|a_k\right|^2\right)-\sum_{k}e^{i\vec{k}\cdot \vec{r}}\frac{N_{k,xy}}{2}\left(a_k+a_{-k}^*\right)\\
    &+ \frac{1}{2}\sum_{k,k'}e^{i\vec{k}\cdot \vec{r}}\left(l_\mathrm{ex}^2 k^2 + N_{k,yy}\right)a_{k'}a_{k'-k}^*\,.
\end{aligned}
\end{equation}

In the above formulas, we have used the following relations for the magnetostatic field~\cite{Suhl1957, Guslienko2011}: 
$\vec{h}_\mathrm{m}\left(\vec{m}_0\right) = -\mathbf{N}\cdot\vec{m}_0$, where $\mathbf{N} = \mathrm{diag}\{N_x,N_y,N_z\}$ is the demagnetising tensor, with $N_x+N_y+N_z = 1$, and $\vec{m}_0$ is the uniform magnetisation mode $(k=0)$, while for $k\neq0$, $\vec{h}_\mathrm{m}\left(\vec{m}_k\right) = -\mathbf{N}_k\cdot\vec{m}_k$, where $N_{k,ij} = (1-s_d(k))k_ik_j/(k^2)$ for $(i,j) \in \{x,y\}$ and $N_{k,zz} = s_d(k)$, with $s_d(k) = (1-e^{-kd})/(kd)$, and $d$ the thickness of the sample.
Notice that for $k\rightarrow 0$, $s_d\rightarrow 1$, meaning that $N_{k,zz}\rightarrow 1$ and $N_{k,ij}\rightarrow 0$.
For the RF field, we have used the following expansion: $h_\mathrm{RF} = \sum_k h_{\mathrm{RF},k}e^{j\vec{k}\cdot \vec{r}}$.
Substituting the Eqs.~(\ref{eq:mpm}, \ref{eq:heffp}, \ref{eq:heffx}) in Eq.~\eqref{eq:LLmp}, and using Fourier orthogonality, we arrive at the following system of equations that describes the nonlinear dynamics of the spin-waves amplitudes $a_k$ with $k\neq0$:
\begin{equation}\label{eq:SWM_k}
\begin{aligned}
 -i\frac{\mathrm{d}a_k}{\mathrm{d}t} =& -h_{\mathrm{RF},k} +\left(H-N_x+l_\mathrm{ex}^2k^2 +\frac{N_{k,yy}+N_{k,zz}}{2}\right) a_k + \frac{N_{k,yy}-N_{k,zz}}{2}a_{-k}^*\\
 &-\sum_{k'} \frac{N_{k',xy}}{2}\left(a_{k'}+a_{-k'}^*\right)a_{k-k'}-\frac{N_{k,xy}}{2}\sum_{k'}a_{k'}a_{k'-k}^*\\
 & +\frac{1}{2}\sum_{k',k''}h_{\mathrm{RF},k''}a_{k'-k''}a_{k'-k}^* + \frac{N_x}{2}\sum_{k'}\left|a_{k'}\right|^2a_k\\
 &-\frac{1}{2}\left(\frac{N_y+N_z}{2}a_0 + \frac{N_y-N_z}{2}a_0^*\right)\sum_{k'} a_{k'}a_{k'-k}^*\\
 &+\frac{1}{2}\sum_{k',k''}(l_\mathrm{ex}^2k'^2 + N_{k',xx}) a_{k''}a_{k''-k'}^*a_{k-k'}\\
 &-\frac{1}{2}\sum_{k',k''}\left[\left(l_\mathrm{ex}^2k'^2 + \frac{N_{k'',yy}+N_{k'',zz}}{2}\right)a_{k''} +\frac{N_{k'',yy}-N_{k'',zz}}{2}a_{-k''}^*\right] a_{k'-k''}a_{k'-k}^*\\
 &+\frac{1}{2}\sum_{k',k'',k} \frac{N_{k'',xy}}{2}a_{k'-k''}a_{k'-k}^*a_{k'''}a_{k'''-k''}^*\,,
\end{aligned}
\end{equation}
and to the following one for the uniform mode $k=0$:
\begin{equation}\label{eq:SWM_0}
\begin{aligned}
 -i\frac{\mathrm{d}a_0}{\mathrm{d}t} =& -h_{\mathrm{RF},0} +\left(H-N_x +\frac{N_{y}+N_{z}}{2}\right) a_0 + \frac{N_{y}-N_{z}}{2}a_{0}^*\\
 &-\sum_{k'} \frac{N_{k',xy}}{2}\left(a_{k'}+a_{-k'}^*\right)a_{-k'}\\
 & +\frac{1}{2}\sum_{k',k''}h_{\mathrm{RF},k''}a_{k'-k''}a_{k'}^*\\
 &-\frac{1}{2}\left[\left(\frac{N_y+N_z}{2}-N_x\right)a_0 + \frac{N_y-N_z}{2}a_0^*\right]\sum_{k'} \left|a_{k'}\right|^2\\
 &-\frac{1}{2}\sum_{k',k''}\left[\left(\frac{N_{k'',yy}+N_{k'',zz}}{2}-N_{k',xx}\right)a_{k''} +\frac{N_{k'',yy}-N_{k'',zz}}{2}a_{-k''}^*\right] a_{k'-k''}a_{k'}^*\\
 &+\frac{1}{2}\sum_{k',k'',k} \frac{N_{k'',xy}}{2}a_{k'-k''}a_{k'}^*a_{k'''}a_{k'''-k''}^*\,.
\end{aligned}
\end{equation}
All the coefficients of Eq.~\eqref{eq:SWM_k} possess a symmetry with respect to he inversion of $k$. 
Then, if exist, the steady states of $a_k$ and $a_{-k}$ will be the same, or if $a_k(0) = a_{-k}(0)$, then $a_k(t) = a_{-k}(t)$ for $t\geq0$.

The use of the \ac{SWM} for numerical computations does not give any advantage with respect to standard micromagnetics. 
The numerical complexity grows with $\mathcal{O}\left(N\left(\log N\right)^2\right)$ due to the presence of terms with double convolution, while for micromagnetics solvers~\cite{MaGICo} it grows with $\mathcal{O}\left(N\log N\right)$, where $N$ is the number of discretisation cells of the $k$-space for the \ac{SWM} and of the real space for the micromagnetic code respectively. 
However, in the next section and in the main text too, we show that, by using general arguments, it is possible to extrapolate from the \ac{SWM} simplified models that allow us to describe with analytical methods the linear dynamics and the parametric instability of the spin-wave amplitudes.

\subsection{Linear Dynamics}

Let us start considering the linear and conservative dynamics of spin wave amplitudes. 
In this respect, we take from Eqs.~\eqref{eq:SWM_k} and \eqref{eq:SWM_0} only the linear terms and do not consider the \ac{RF}-field term. Then, we have the following equation:
\begin{equation}\label{eq:SWM_lin_cons}
    -i\frac{\mathrm{d}a_k}{\mathrm{d}t} = A_k a_k + B_k a_{-k}^*\,,
\end{equation}
where the expressions for $A_k$ and $B_k$ as a function of the material parameters and the $k$ index, for every $k$, can be directly obtained from the Eqs.~\eqref{eq:SWM_k} and \eqref{eq:SWM_0}. 
The above equation has a Hamiltonian structure but is not in a diagonal form. 
In order to preserve the structure and diagonalise it, we perform the following change of variables~\cite{Suhl1957, Zakharov1975}:
\begin{equation}\label{eq:Bog_transf}
\begin{aligned}
    b_k & = \varepsilon_k a_k +\eta_k a_{-k}^*\,,\\
    \varepsilon_k & = \cosh\frac{\psi_k}{2}\,, \\
    \eta_k & = \sinh\frac{\psi_k}{2}\,,\\
    \tanh\psi_k & = \frac{B_k}{A_k}\,.
\end{aligned}  
\end{equation}
 According to this transformation, Eq.~\eqref{eq:SWM_lin_cons} takes the following form:
 \begin{equation}\label{eq:SWNM_lin_cons}
 -i\frac{\mathrm{d}b_k}{\mathrm{d}t} = \varepsilon_k\left(-i\frac{\mathrm{d}a_k}{\mathrm{d}t}\right) -\eta_k\left(i\frac{\mathrm{d}a_{-k}^*}{\mathrm{d}t}\right) = \omega_k b_k\,,
\end{equation}
where the relations: 
 \begin{equation}\label{eq:disp_rel}
\omega_k = \sqrt{A_k^2-B_k^2} =
\begin{cases}
&\sqrt{\left(H-N_x+l_\mathrm{ex}^2k^2 +N_{k,yy}\right)\left(H-N_x+l_\mathrm{ex}^2k^2 +N_{k,zz}\right)}\,,\\
&\sqrt{\left(H-N_x+N_y\right)\left(H-N_x+N_z\right)}\,,
\end{cases}
 \end{equation}
are in the order from top to bottom, the dispersion relation for $k\neq0$ and Kittel's frequency $k = 0$. 
The variables $b_k$ will be called spin-wave normal modes because their conservative dynamic is governed by the equations of a set of independent harmonic oscillators.

The inclusion of damping effects in the model follows the same procedure used to derive terms due to the conservative torque of the Landau-Lifshitz equation. 
In this respect, one can easily show that the linear dynamics, including damping effects and the \ac{RF}-field term, is described by the following equation:
\begin{equation}\label{eq:SWM_lin_damp_rf}
    -i\frac{\mathrm{d}a_k}{\mathrm{d}t} = -h_{\mathrm{RF},k} + (1+i\alpha)(A_k a_k + B_k a_{-k}^*)\,.
\end{equation}
Then, if we rewrite it in terms of the spin-wave normal modes, we get the following equation:
\begin{equation}\label{eq:NSWM_lin_damp_rf}
    -i\frac{\mathrm{d}b_k}{\mathrm{d}t} = -h_{k} + \omega_k b_k +i\alpha(A_k b_k + B_k b_{-k}^*)\,,
\end{equation}
where $h_k = \varepsilon_k h_{\mathrm{RF},k} -\eta_k h_{\mathrm{RF},-k}^*$. 
The amplitude of the \ac{RF}-field is a sinusoidal function in time, therefore, it can be decomposed into two circular polarised fields: one with clockwise rotation in the complex plane $h_{k}^+$ and the other with counter-clockwise rotation $h_{k}^-$. 
This means that each $b_k$ has both components $(b_k^+,\,b_k^-)$ too. 
We are interested in the spin-wave normal mode amplitudes close to the resonance condition $\omega_k\approx\omega_\mathrm{RF}$. 
Then, with good approximation, the relation $b_k\approx b_k^+$ can be assumed. 
When Eq.~\eqref{eq:NSWM_lin_damp_rf} is written in a rotating complex plane with angular frequency $\omega_\mathrm{RF}$ and only slow varying terms are kept, by using the phasor notation $b_k(t) = \overline{b}_k\, \exp{(i\omega_\mathrm{RF}t)}$, we have the following relation:
\begin{equation}\label{eq:lin_kth_sol}
    \overline{b}_k = \frac{-\overline{h}_k^+}{\left(\omega_\mathrm{RF}-\omega_k\right)-i\alpha A_k}\,.
\end{equation}
The symmetry $a_k(t) = a_{-k}(t)$ discussed in the previous section implies the same symmetry on the $b_k$ independently of the dynamics (as a special case, the same property holds for Eq.~\eqref{eq:lin_kth_sol}). 
This is used in the main text for the derivation of the threshold \ac{RF} field for the spin-wave parametric instability.

\subsection{Spin-Wave Parametric Instability}
At low excitation power, the spin-wave dynamics remain linear, and only Damon-Eshbach (DE) spin waves and the uniform mode are excited. 
Their amplitudes can be calculated using the linearised equations of motion, involving the spin-wave amplitudes. 

In the following, we assume that the RF frequency is such that the uniform mode is off-resonance, so it is not significantly excited. 
Independent of the RF frequency, MMM images reveal that beyond a certain RF power, additional spin waves are excited, despite not coupling directly to the RF field.
These additional modes exhibit wave vectors in directions different from that of the RF excitation, which is a hallmark of parametric instability.

To explain this, two types of nonlinear magnon scattering processes are considered:
\begin{enumerate}
    \item \emph{Three-magnon scattering process}: The \ac{SWM} (see Eq.~\eqref{eq:SWM_k} and Eq.~\eqref{eq:SWM_0}) involves a summation of terms, each of the following form: $h_{\mathrm{RF},k''}a_{k'-k''}a_{k'-k}^*$. 
    The relevant contributions are those where a parametrically excited mode couples with both an RF-driven DE spin wave mode and the RF field. 
    This restricts the summation to terms with $k' = 0$ or $k'= k''+ k$.
    \item \emph{Four-magnon scattering process: } 
    This process involves two DE spin waves and two parametrically excited spin waves with opposite wave vectors ($\pm k$).
    The relevant terms are of the form:  $ a_{k'-k''}a_{k''}a_{k'-k}^*$ (see Eq.~\eqref{eq:SWM_k}). 
    Restricting again to $k'=0$ or $k' = k''+k$, we identify interactions where DE spin waves  $a_{\pm k''}$ scatter into parametrically, not directly RF-field-excited modes $a_{\pm k}$. 
    This leads to the following simplified \ac{SWM}: 
    \begin{equation}\label{eq:nonlin_param_ak}
    \begin{aligned}
        -i\frac{\mathrm{d}a_k}{\mathrm{d}t} = & \left(1+i\alpha\right)\left(A_k a_k + B_k a_{-k}^*\right)+\frac{1}{2}\sum_{k'\in \{k_\mathrm{DE}\}}h_{\mathrm{RF},k'}\left(a_{k'}^*a_k+a_{-k'}a_{-k}^*\right)\\
        & -\frac{1}{2}\sum_{k'\in \{k_\mathrm{DE}\}}\left(C_{k'}\left|a_{k'}\right|^2+B_{k'}a_{k'}^*a_{-k'}^*\right)a_k\\
        & -\frac{1}{2}\sum_{k'\in \{k_\mathrm{DE}\}}\left|a_{k'}\right|^2 \left(C_k a_k + B_k a_{-k}^*\right)\,,
    \end{aligned}
    \end{equation}
    where 
    \begin{equation}
    \begin{aligned}
        A_k & = H-N_x+l_\mathrm{ex}^2k^2 + (N_{k,yy}+N_{k,zz})/2 \, ,\\
        B_k & = (N_{k,yy}-N_{k,zz})/2 \, ,\\
        C_k & = l_\mathrm{ex}^2k^2 + (N_{k,yy}+N_{k,zz})/2 \, . 
    \end{aligned} \nonumber
    \end{equation}  
    Here, the coefficients $A_k$, $B_k$, $C_k$ describe dipolar and exchange contributions, and the set $\{k_\mathrm{DE}\}$ includes the \ac{DE} spin wave modes. 
\end{enumerate}

\paragraph{Parametric Equations}
In the following, the derivation of the parametric terms that appear in the equations of spin-wave normal modes is given.
The starting point is Eq.~\eqref{eq:nonlin_param_ak}. 
Each term on the right-hand side transforms according to Eq.~\eqref{eq:SWNM_lin_cons}. 
Let us consider the nonlinear terms due to the RF-field. 
The following transformation relation results:
\begin{equation}\label{eq:nonlin_term_hrf}
\begin{aligned}
h_{\mathrm{RF},k'}(a_{k'}^*a_k+a_{-k'}a_{-k}^*)  \rightarrow & \, \left(\varepsilon_k h_{\mathrm{RF},k'}-\eta_k h_{\mathrm{RF},-k'}^*\right)\left(a_{k'}^*a_k+a_{-k'}a_{-k}^*\right)\,\\ 
 \approx & \, \frac{1}{\omega_k\omega_{k'}}\left[\left(A_k A_{k'}+B_k B_{k'}+\omega_k\omega_k'\right)h_{k'}^+ \right. \\
         & \quad \quad \quad \quad \left.- \left(A_k B_{k'}+B_k A_{k'}\right)h_{-k'}^{-*}\right]b_{-k'}b_{-k}^*\,,
\end{aligned}
\end{equation}
where the approximated equality symbol means that only parametric terms have been considered.
The last equation represents the first parametric term that augments the linear model of spin-wave normal mode dynamics and will allow us to investigate the threshold of parametric instability.
The nonlinear term describing the four-magnon scattering process in the Eq.~\eqref{eq:nonlin_param_ak}, from which the second parametric term is derived, transforms according to the following relation:
\begin{equation}\label{eq:param_term_4magn_1}
\begin{aligned}
    \MoveEqLeft[3] \left|a_{k'}\right|^2\left[\left(C_k+C_{k'}\right)a_k + B_k a_{-k}^*\right]+B_{k'}a_{k'}^*a_{-k'}^*a_k\,\\
    \rightarrow \quad  & \left|a_{k'}\right|^2 \left[\left(C_k+C_{k'}\right)\left(\varepsilon_k a_k-\eta_k a_{-k}^*\right)+B_k\left(\varepsilon_ka_{-k}^*-\eta_ka_k\right)\right]\\
     & + B_{k'}(a_{k'}^*a_{-k'}^*\varepsilon_{k}a_k-a_{-k'}a_{k'}\eta_k a_{-k}^*)\\
     \approx \quad &  \frac{B_kB_{k'}}{2\omega_k\omega_{k'}}\left[\left(C_k+C_{k'}-A_k\right)b_{k'}^2-A_{k'}b_{k'}b_{-k'}\right]b_{-k}^*\,.
\end{aligned}
\end{equation}
When both parametric terms are inserted into the equation of dynamics for $b_k(t)$, and $b_{k'}$ is expressed according to Eq.~\eqref{eq:lin_kth_sol}, one obtains the following equation:
\begin{equation}\label{eq:nonlin_param_bk_comp}
\begin{aligned}
   -i\frac{\mathrm{d}b_k}{\mathrm{d}t} = & (\omega_k + i\alpha A_k) b_k \\
  & + \sum_{k'\in \{k_\mathrm{DE}\}}\left(\frac{\left(\chi_{k,k'}^+h_{k'}^++\chi_{k,k'}^-h_{k'}^{-*}\right)h_{k'}^+}{\omega_\mathrm{RF}-\omega_{k'}-i\alpha A_{k'}}+\frac{\zeta_{k,k'}h_{k'}^{+2}}{(\omega_\mathrm{RF}-\omega_{k'}-i\alpha A_{k'})^2}\right)\,b_{-k}^*\,,
   \end{aligned}
\end{equation}
where:
\begin{equation}
    \begin{aligned}
       & \chi_{k,k'}^+=\frac{1}{2\omega_k\omega_{k'}}\left(A_kA_{k'}+B_kB_{k'}+\omega_k\omega_k'\right)\,,\\
       & \chi_{k,k'}^-=-\frac{1}{2\omega_k\omega_{k'}}\left(A_kB_{k'}+B_kA_{k'}\right)\,,\\
       & \zeta_{k,k'} = -\frac{B_kB_{k'}}{2\omega_k\omega_{k'}}\left(H-N_x\right)\,.
    \end{aligned}
\end{equation}
The linear resonance condition of the \ac{DE} modes for $\omega_{k'}\approx\omega_\mathrm{RF}$ maximises the magnitude of the coefficient of the parametric terms and then minimises the threshold for the parametric instability~\cite{Suhl1957}. 
Moreover, in such a condition, these two types of coefficient scale differently with respect to $\alpha$. 
For the \ac{YIG} medium, the value of the damping constant $\alpha\approx 10^{-4}$ is such that the coefficient of the parametric term due to the four-magnon scattering process for $\omega_{k'}\approx\omega_\mathrm{RF}$ is several orders of magnitude larger than the coefficient of the three-magnon scattering mediated by the \ac{RF} field. 
Therefore, this last one can be safely neglected.

\paragraph{Critical Field}

Before the instability sets in, we assume DE spin waves oscillate in a steady linear state (see Eq.~\eqref{eq:lin_kth_sol}). 
Transforming Eq.~\eqref{eq:nonlin_param_ak} into the normal-mode basis $b_k$ , and focusing on four-magnon parametric terms, we arrive at:
\begin{equation}\label{eq:nonlin_param_bk}
   -i\frac{\mathrm{d}b_k}{\mathrm{d}t} = \left(\omega_k + i\alpha A_k\right) b_k + \sum_{k'\in \{k_\mathrm{DE}\}}\frac{\zeta_{k,k'} h_{k'}^{+2}}{\left(\omega_\mathrm{RF}-\omega_{k'}-i\alpha A_{k'}\right)^2}\,b_{-k}^*\,.
\end{equation}
This equation represents a parametric oscillator, where the second term triggers instability by coupling $b_{k}$ to the conjugate mode $b_{-k}^*$. 
Following Suhl’s approach~\cite{Suhl1957}, the critical RF field for instability is:
\begin{equation}\label{eq:thresh_field_inst_1}
    h_\mathrm{RF,crit} = \left(\frac{\left(\omega_\mathrm{RF}-\omega_k\right)^2+\alpha^2A_k^2}{\left|\sum_{k'\in \{k_\mathrm{DE}\}}\frac{\zeta_{k,k'}\hat{h}_{k'}^{+2}}{\left(\omega_\mathrm{RF}-\omega_{k'}-i\alpha A_{k'}\right)^2}\right|^2}\right)^\frac{1}{4}\,,
\end{equation}
where we have used the following relation: $\overline{h}_{k'}^+ = h_\mathrm{RF}\,\hat{h}_{k'}^+$, with $h_\mathrm{RF}$ the amplitude of the \ac{RF} field.

\paragraph{Inclusion of the Uniform Mode Dynamics}

\begin{figure}[bth!]
  \centering  %
  \includegraphics[width=0.576\textwidth]{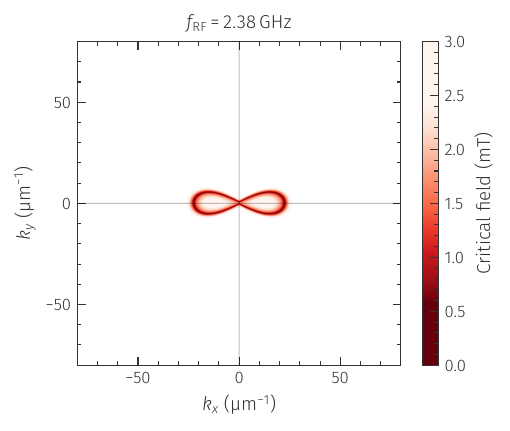} 
  \caption[ ]{Theoretical threshold field distribution in $k$-space for a frequency of $f_\mathrm{RF} = \qty{2.38}{GHz}$}
  \label{fig:hcrit_k_f}
\end{figure}

When the RF-field frequency approaches the uniform (Kittel) mode, its contribution can no longer be ignored. Following Suhl~\cite{Suhl1957}, additional parametric terms of the type $b_0^2b_{-k}^*$ must be added to Eq.~\eqref{eq:nonlin_param_bk}. 
Such terms are obtained via the transformation terms of the \ac{SWM} given by the product of one among $a_0^2,\,{a_0^*}^2,\,\left|a_0\right|^2$ with one of the terms $a_k,\, a_{-k}^*$.
Interestingly, Eqs.~\eqref{eq:nonlin_param_bk} and \eqref{eq:thresh_field_inst_1} still hold but $k'\in\{0,k_\mathrm{DE}\}$, where 
\begin{equation}
        \zeta_{k,0} = B_0\frac{A_k\left(B_0+B_k\right) + B_k\left(A_0-\omega_0\right) -B_k D_k}{4\omega_0\omega_k} - \frac{\left(A_0+\omega_0\right)A_kE_k}{4\omega_0\omega_k}
\end{equation}
with 
\begin{equation}
\begin{aligned}
A_0 & = H-N_x + (N_y+N_z)/2\\
B_0 & = (N_y-N_z)/2\\
D_k &= (N_y+N_z)/2-N_x + (N_{k,yy}+N_{k,zz})/2-N_{k,xx}\\
E_k &= (N_y + N_z)/2-l_\mathrm{ex}^2k^2-N_{k,xx} \ .
\end{aligned} \nonumber
\end{equation} 
 In Fig.~\ref{fig:nonlinear_exp_theo}c and in Fig.~\ref{fig:hcrit_k_f}, we show the threshold field distribution in $k$-space for two \ac{RF} frequencies according to the theory discussed in the Methods section of the main text. 
Given the \ac{RF} frequency, the threshold field values are significantly lower on the dispersion curve in the $k$-space corresponding to $\omega_k = \omega_\mathrm{RF}$, and, as expected, the higher the efficiency of the \ac{CPW}, the lower the critical field value. 

\paragraph{Connection to Suhl's Instability}

In materials like YIG with low damping $\alpha\approx 10^{-4}$, only wave vectors $k'$ with $\omega_{k'} \approx \omega_\mathrm{RF}$, such that the CPW coupling efficiency is not zero, contribute significantly to the denominator in Eq.~\eqref{eq:thresh_field_inst_1}. 
We define $\tilde{k}$ the $k$-space indices such that this relation is satisfied. 
Then, the sum over the interval $\{0,k_\mathrm{DE}\}$ can be replaced in good approximation by the sum over $\{\tilde{k}\}$. 
In this respect, when the CPW efficiency peaks at $\{\tilde{k}\}$, Eq.~\eqref{eq:thresh_field_inst_1} becomes:
\begin{equation}\label{eq:thresh_field_inst_4}
    h_\mathrm{RF,crit} \approx \frac{\left[\left(\omega_\mathrm{RF}-\omega_k\right)^2+\alpha^2A_k^2\right]^\frac{1}{4}\left[\left(\omega_\mathrm{RF}-\omega_{\tilde{k}}\right)^2+\alpha^2A_{\tilde{k}}^2\right]^\frac{1}{2}}{\left(2\left|\zeta_{k,\tilde{k}}\hat{h}_{\tilde{k}}^{+2}\right|\right)^\frac{1}{2}}\,.
\end{equation}
This equation reduces to Suhl’s expression for the second-order instability~\cite{Suhl1957} for $\tilde{k}=0$, and for the CPW coupling efficiency peaked at $k=0$. 
Therefore, the spin-wave instability described here represents a generalised form of Suhl's second-order instability in which the modes responsible for the parametric excitation are, in principle, all those that directly couple to the \ac{RF} field due to the coupling efficiency of the \ac{CPW}.  
In our case, these modes are a subset of the \ac{DE} spin waves.

\paragraph{Threshold Field for Parametric Excitation of Spin Waves}

The threshold field for spin wave parametric excitation corresponds to the minimum in the $k$-space of the critical field expressed by Eq.~\eqref{eq:thresh_field_inst_4}.
The minimization of the critical field can be done in two steps. 
The first one corresponds to set the resonance condition $\omega_k=\omega_{\tilde{k}}=\omega_\mathrm{RF}$ in the above relation that can be written as:
\begin{equation}\label{eq:thresh_field_inst_5}
    h_\mathrm{RF,crit} \approx \sqrt{\frac{A_k A_{\tilde{k}}}{\left|B_k B_{\tilde{k}}\right|}}\,\frac{\alpha^{3/2}\omega_\mathrm{RF} \sqrt{A_{\tilde{k}}}}{\left|\hat{h}_{\tilde{k}}^+\right|\sqrt{H-N_x}}\,.
\end{equation}
The second step corresponds to select among the infinite spin wave modes that satisfy the resonance condition, those that minimize the factor $A_k/|B_k|$. 
This factor takes into account the elliptical polarization of the spin wave amplitudes $a_k$ and in the limiting case where it becomes circular, $B_k\rightarrow0 \Rightarrow h_\mathrm{RF,crit}\rightarrow\infty$. 
Then, the occurrence of the spin wave parametric instability is forbidden when the magnetic equilibrium is uniform and out of the film plane ($z$-direction).
The minimum of the factor $A_k/|B_k|$ depends on the \ac{RF} field frequency. For frequency values considered in the experiments, we found that it is minimum when $k = \tilde{k}$, which corresponds to the \ac{DE} spin-wave modes such that $\omega_{\tilde{k}}=\omega_\mathrm{RF}$.
In this respect, Eq.~\eqref{eq:thresh_field_inst_5} becomes
\begin{equation}\label{eq:thresh_field_inst_6}
    h_\mathrm{RF,thr} = \min_{k}h_\mathrm{RF,crit} \approx \frac{A_{\tilde{k}}}{\left|B_{\tilde{k}}\right|}\,\frac{\alpha^{3/2}\omega_\mathrm{RF} \sqrt{A_{\tilde{k}}}}{\left|\hat{h}_{\tilde{k}}^+\right|\sqrt{H-N_x}}\,.
\end{equation}
When the resonance condition is not satisfied any more, the critical field increases rapidly, namely, the parametric excitation occurs only for spin-wave modes in a narrow region in the $k$-space, where the resonance condition $\omega_k\approx\omega_{\tilde{k}} \approx \omega_\mathrm{RF}$ is satisfied.\\

\paragraph{Magnetization Deflection-Angle at the Instability Condition}
The excitation of certain spin wave modes produces a deflection-angle between the local magnetization vector and the direction of the magnetic equilibrium at each point of the film. 
Once the magnetization component perpendicular to the equilibrium direction $(m_\perp)$ is known, such angle can be determined by the following relation:
\begin{equation}
\theta(\vec{r},t) = \arctan{\left(\frac{m_\perp}{\sqrt{1-m_\perp^2}}\right)}\Bigg|_{m_\perp\ll1}\approx m_\perp\,,
\end{equation}
where according to Eqs.~\eqref{eq:mpm} and \eqref{eq:Bog_transf}, one has
\begin{equation}\label{eq:mperp}
m_\perp = |m_\pm| = \sqrt{\sum_{k,k'} (\varepsilon_{k'} b_{k'} -\eta_{k'} b_{-k'}^*)(\varepsilon_{k''} b_{-k''}^* -\eta_{k''} b_{k''}) e^{j(\vec{k}'+\vec{k}'')\cdot \vec{r}}}\,.
\end{equation}
Prior to the spin-wave parametric instability, the only spin-wave modes excited are those $\tilde{k}\in\{k_\mathrm{DE}\}:\,\omega(\tilde{k}) = \omega_\mathrm{RF}$, namely, spin waves resonant with the RF field. 
The spin wave dynamics before the instability is assumed linear, and the steady oscillations are described by Eq.~\eqref{eq:lin_kth_sol}. 
By considering it and the following property of the spin-wave normal-mode amplitudes: $b_k(t) = b_{-k}(t)$, Eq.~\eqref{eq:mperp} reduces to 
\begin{equation}\label{eq:mperp2}
m_\perp = \sqrt{\left[\left(\varepsilon_{\tilde{k}}^2+\eta_{\tilde{k}}^2\right)\left|b_{\tilde{k}}\right|^2-\varepsilon_{\tilde{k}}\eta_{\tilde{k}}\left(b_{\tilde{k}}^2+b_{\tilde{k}}^{*2}\right)\right]}\left|\cos{\tilde{k}y}\right|\,.
\end{equation} 
The maximum of the deflection angle corresponds to the maximum of the perpendicular component of the magnetization field. Then, if the above relation is maximized with respect to the point and time instant, one can arrive to the following relation:
\begin{equation}
m_{\perp}  = \frac{h_\mathrm{RF}|\hat{h}_{\tilde{k}}^+|}{\alpha\sqrt{A_{\tilde{k}}\omega_\mathrm{RF}}} \,.
\end{equation}
At the instability condition, the \ac{RF} field value is given by Eq.~\eqref{eq:thresh_field_inst_5}, and therefore the perpendicular magnetization component can be expressed as:
\begin{equation}
m_{\perp,\max}  \approx \frac{A_{\tilde{k}}}{\left|B_{\tilde{k}}\right|}\sqrt{\frac{\alpha\,\omega_\mathrm{RF}}{H-N_x}} \,.
\end{equation}
In the case $f_\mathrm{RF} = \qty{2.38}{GHz}$, $m_{\perp,\max}\approx 0.03$ that corresponds to a maximum magnetization deflection-angle of $\theta_\mathrm{max}\approx \ang{2}$, while for $f_\mathrm{RF} = \qty{9.00}{GHz}$, $m_{\perp,\max}\approx 0.15$ and $\theta_\mathrm{max}\approx \ang{10}$. 
Such values are consistent with the assumption of linearity that has been used for the derivation of the threshold RF field value. 
Indeed, it corresponds to small tilting angles of the local magnetization with respect to the equilibrium direction.

\end{appendices}

\newpage

\bibliography{bibliography}

\end{document}